\documentclass[aps,prd,twocolumn,showpacs,nofootinbib]{revtex4-1}

\usepackage{graphicx}
\usepackage{amsmath}
\usepackage{amssymb}
\usepackage{bm}
\usepackage{bbm}
\usepackage{color}
\usepackage{slashed}
\usepackage[colorlinks=true,linkcolor=blue,citecolor=blue, urlcolor=blue]{hyperref}

\begin{document}

\title{Next-to-leading order QCD corrections to $Z\to \eta_Q+Q+\bar{Q}$}

\author{Xu-Chang Zheng$^{a}$}
\email{zhengxc@cqu.edu.cn}
\author{Xing-Gang Wu$^{a}$}
\email{wuxg@cqu.edu.cn}
\author{Xi-Jie Zhan$^{a}$}
\email{zhanxj@cqu.edu.cn}
\author{Hua Zhou$^{a,b}$}
\email{zhouhua@cqu.edu.cn}
\author{Hong-Tai Li$^{a}$}
\email{liht@cqu.edu.cn}

\affiliation{$^a$ Department of Physics, Chongqing Key Laboratory for Strongly Coupled Physics, Chongqing University, Chongqing 401331, People's Republic of China\\
$^b$ Department of Physics,
Norwegian University of Science and Technology, H{\o}gskoleringen 5, N-7491 Trondheim, Norway}

\begin{abstract}

It has been found that at a high luminosity $e^+ e^-$ collider, sizable $\eta_c+c\bar{c}X$ and $\eta_b+b\bar{b}X$ events can be produced when it works around the $Z$ peak. In this paper, we calculate the decay widths of $Z \to \eta_c+c+\bar{c}+X$ and $Z \to \eta_b+b+\bar{b}+X$ up to next-to-leading order (NLO) accuracy. We find that the NLO corrections are significant in these two processes. After including the NLO corrections, the decay widths of $Z \to \eta_c+c+\bar{c}+X$ and $Z \to \eta_b+b+\bar{b}+X$ are enhanced by about $34\%$ and $28\%$ for the case of $\mu_R=2m_Q$, and are enhanced by about $112\%$ and $83\%$ for the case of $\mu_R=m_{_Z}$, respectively. The differential decay widths $d\Gamma/dz$, $d\Gamma/dm_{12}$ and $d\Gamma/dm_{23}$ for these two decay processes are also analyzed.

\end{abstract}


\maketitle

\section{Introduction}
\label{secIntro}

Heavy quarkonium production presents an ideal laboratory for the study of the interplay between the perturbative and nonperturbative effects of QCD; it has been a focus of theoretical and experimental interest since the discovery of $J/\psi$ in 1974. In order to describe the quarkonium production, the color-evaporation model (CEM) \cite{Fritzsch:1977ay, Halzen:1977rs}, the color-singlet model (CSM) \cite{Chang:1979nn, Berger:1980ni, Matsui:1986dk}, and the nonrelativistic QCD (NRQCD) effective theory \cite{nrqcd} have been proposed. Among them, the NRQCD effective theory provides a systematic way of separating the short-distance and long-distance effects in the quarkonium production, and has achieved great success in describing the experimental data of the quarkonium production, especially for the unpolarized cross section of the $J/\psi$ hadroproduction \cite{ybook1, Andronic:2015wma, Lansberg:2019adr, Chen:2021tmf}. However, there are still challenges to NRQCD. For instance, the hadroproduction cross section of $\eta_c$ measured by the LHCb experiments \cite{Aaij:2014bga} can be well described by the color-singlet contribution, i.e., the color-octet contribution should be very small \cite{Butenschoen:2014dra}. This seems inconsistent with the heavy-quark spin symmetry (HQSS) relation between the long-distance matrix elements (LDMEs) of $\eta_c$ and $J/\psi$.\footnote{References \cite{Han:2014jya, Zhang:2014ybe} pointed out that the hadroproduction data of $J/\psi$ and $\eta_c$ can be simultaneously described by one set of LDMEs. However, theoretical predictions based on this set of LDMEs fail to describe the $J/\psi$ production data from $e^+e^-$ annihilation at the B-factory \cite{Belle:2009bxr, Li:2014fya}.} It is important to study more processes involving the charmonium for testing the NRQCD factorization.

It has been found that the heavy quarkonium production through $Z$ boson decays can provide a good platform for studying the production mechanism of quarkonia, which has attracted great attention \cite{Guberina:1980dc, Keung:1980ev, Abraham:1989ri, Barger:1989cq, Hagiwara:1991mt, Braaten:1993mp, Fleming:1993fq, Cheung:1995ka, Cho:1995vv, Ernstrom:1996aa, Schuler:1997is, JXWang, Liao:2015vqa, Huang:2014cxa, Likhoded:2017jmx, Bodwin:2017pzj, Sun:2018hpb, Chung:2019ota, jpsiFFNLO, Sun:2020yrb, Sun:2021avu, Zheng:2021jyd, Sun:2022iir}. A large number of $Z$ boson events can be accumulated at the LHC or a future high-luminosity $e^+e^-$ collider running around the $Z$ pole. At the LHC, there are about $10^9$ $Z$ bosons to be produced per year \cite{Liao:2015vqa}. It is well known that several proposed high-luminosity $e^+e^-$ colliders, such as the  ILC \cite{Baer:2013cma}, FCC-ee \cite{FCC:2018byv}, CEPC \cite{CEPCStudyGroup:2018ghi}, and Super $Z$ Factory \cite{zfactory}, are planned to run at the $Z$ pole for a period of time. When the $e^+e^-$ collider runs at the $Z$ pole and with a luminosity of $10^{34- 36}{\rm cm}^{-2}{\rm s}^{-1}$, there are about $10^{9-11}$ $Z$ bosons to be produced per year \cite{Erler:2000jg}. These colliders will open new opportunities for studying the quarkonium production through $Z$ boson decays.

Most studies on the heavy quarkonium production through the $Z$ boson decays focus on the spin-triplet $J/\psi$ and $\Upsilon$ production, while few studies are for the spin-singlet $\eta_Q$ ($Q=b$ or $c$) production. In our recent work \cite{Zheng:2021jyd}, we studied the inclusive production of the $\eta_Q$ via the $Z$ boson decays up to order $\alpha \alpha_s^2$ within the framework of NRQCD, in which the leading color-singlet ($^1S_0^{[1]}$) and color-octet ($^1S_0^{[8]}$, $^3S_1^{[8]}$, and $^1P_1^{[8]}$) Fock states are considered. The study found that there are many interesting features in these production processes. An important channel contributing to the inclusive production $Z\to \eta_Q+X$ is $Z \to \eta_Q+Q+\bar{Q}$. Experimentally, its decay width can be measured separately through the heavy-flavor tagging technology. Therefore, it is helpful to do a precise theoretical study on this channel. In this paper, we devote ourselves to studying the decay $Z \to \eta_Q+Q+\bar{Q}+X$, which starts at order $\alpha \alpha_s^2$, up to NLO QCD accuracy. We will use the CSM, which is the leading-order (LO) contribution (in $v_Q$, where $v_Q$ is the velocity of the heavy quark or the heavy antiquark in the quarkonium rest frame, $v_c^2 \approx 30\%$ for the $\eta_c$ and $v_b^2 \approx 10\%$ for the $\eta_b$ \cite{Buchmuller:1980su}) of NRQCD,\footnote{The next-order relativistic correction to the color-singlet contribution is suppressed by order $v_Q^2$, while the color-octet contribution is suppressed by order $v_Q^4$. It is noted that the short-distance factor of the color-octet contribution may be enhanced compared to that of the color-singlet contribution. In this work, we assume the color-octet contribution is very small, and focus on the color-singlet contribution.} to calculate the decay width of $Z \to \eta_Q+Q+\bar{Q}+X$.

The NLO QCD corrections to $Z\to \eta_Q+gg$ have recently been finished through the CSM \cite{Sun:2022iir}. The authors there found that the NLO corrections are significant due to the fragmentation diagrams appearing at the NLO level.\footnote{The large fragmentation contribution in the NLO corrections of $Z\to \eta_Q+gg$ can be calculated through the fragmentation-function approach, where the large logarithms of $m_{_Z}/m_Q$ in higher-order corrections can be resummed through the Dokshitzer-Gribov-Lipatov-Altarelli-Parisi (DGLAP) evolution of the fragmentation functions \cite{Zheng:2021mqr}.} Reference \cite{Sun:2022iir} and the present paper give a complete study on the $\eta_Q$ production through $Z$ boson decays up to NLO QCD accuracy under the CSM.

The remaining parts of the paper are organized as follows. In Sec.\ref{widthLO}, we briefly present useful formulas for the process $Z\to \eta_Q +Q+\bar{Q}+X$ at the LO accuracy. In Sec.\ref{secNLO}, we present the formulas for calculating the NLO QCD corrections to the process $Z\to \eta_Q +Q+\bar{Q}+X$. In Sec.\ref{secNumer}, numerical results and discussions are presented. Section \ref{secSum} is reserved as a summary.

\section{LO decay width}
\label{widthLO}

Under the NRQCD factorization, the decay width for $Z \to \eta_Q+ Q+\bar{Q}+X$ can be written as
\begin{eqnarray}
&& d\Gamma_{Z\to \eta_Q+Q+\bar{Q}+X}\nonumber \\
&& =\sum_n d\tilde{\Gamma}_{Z\to (Q\bar{Q})[n]+Q+\bar{Q}+X}\langle {\cal O}^{\eta_Q}(n)\rangle,\label{nrqcdfact}
\end{eqnarray}
where $d\tilde{\Gamma}$ are short-distance coefficients (SDCs) and $\langle {\cal O}^{\eta_c}(n)\rangle$ are LDMEs. The sum extends over all of the intermediate color-singlet and color-octet states $^{2S+1}L_{J}^{[1,8]}$. In the lowest-order nonrelativistic approximation (i.e., the CSM), only the color-singlet contribution with $n=\, ^1S_0^{[1]}$ needs to be considered.

In the practical calculation, we first calculate the decay width for a free on-shell $(Q\bar{Q})$ pair with quantum number $^1S_0^{[1]}$, i.e., $d\Gamma_{Z\to (Q\bar{Q})[^1S_0^{[1]}]+Q+\bar{Q}+X}$. Then the decay width for the $\eta_Q$ meson can be obtained from $d\Gamma_{Z\to (Q\bar{Q})[^1S_0^{[1]}]+Q+\bar{Q}+X}$ through replacing $\langle {\cal O}^{(Q\bar{Q})[^1S_0^{[1]}]}(^1S_0^{[1]})\rangle$ by $\langle {\cal O}^{\eta_Q}(^1S_0^{[1]})\rangle$.

\begin{figure}[htbp]
\includegraphics[width=0.45\textwidth]{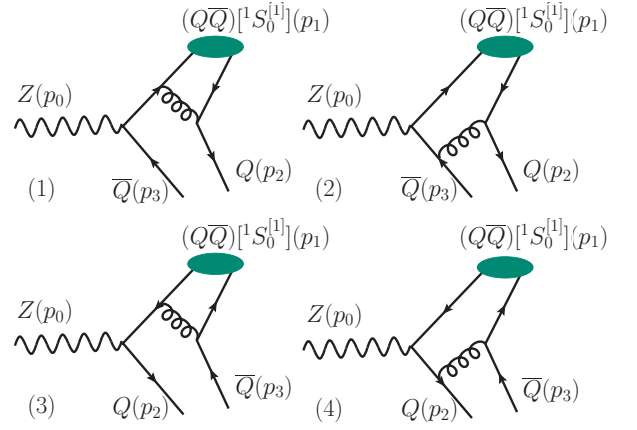}
\caption{The LO Feynman diagrams for $Z \to (Q\bar{Q})[^1S_0^{[1]}]+Q+\bar{Q}$.
 } \label{feylo}
\end{figure}

At the LO level, there are four Feynman diagrams for the decay process $Z \to (Q\bar{Q})[^1S_0^{[1]}]+ Q+\bar{Q}$, which are shown in Fig.\ref{feylo}. Corresponding to the four Feynman diagrams, the LO amplitude for this process can be written as $M_{\rm LO}=M_1+M_2+M_3+M_4$, where
\begin{eqnarray}
iM_1=&&-\frac{i g}{2{\rm cos}\, \theta_W}\frac{-i}{(p_1/2+p_2)^2+i\epsilon}\bar{u}(p_2)(ig_s \gamma^{\mu}T^a) \nonumber \\
&& \cdot \Pi_1 \Lambda_1 (ig_s \gamma_{\mu}T^a) \frac{i}{\slashed{p}_1+\slashed{p}_2-m_Q+i \epsilon} \epsilon_{\nu}(p_0) \gamma^{\nu} \nonumber \\
&& \cdot (V_Q-A_Q\gamma_5)v(p_3), \\
iM_2=&&-\frac{i g }{2{\rm cos}\, \theta_W}\frac{-i}{(p_1/2+p_2)^2+i\epsilon}\bar{u}(p_2)(ig_s \gamma^{\mu}T^a) \Pi_1 \nonumber \\
&&  \cdot \Lambda_1 \epsilon_{\nu}(p_0)\gamma^{\nu} (V_Q-A_Q\gamma_5)\frac{i}{-\slashed{p}_0+\slashed{p}_1/2-m_Q+i \epsilon} \nonumber \\
&&.(ig_s \gamma_{\mu}T^a)v(p_3),\\
iM_3=&&-\frac{i g }{2{\rm cos}\, \theta_W}\frac{-i}{(p_1/2+p_3)^2+i\epsilon}\bar{u}(p_2)\epsilon_{\nu}(p_0)\gamma^{\nu} \nonumber \\
&&  \cdot (V_Q-A_Q\gamma_5) \frac{i}{-\slashed{p}_1-\slashed{p}_3-m_Q+i \epsilon} (ig_s \gamma_{\mu}T^a) \nonumber \\
&&.\Pi_1 \Lambda_1 (ig_s \gamma^{\mu}T^a) v(p_3),\\
iM_4=&&-\frac{i g }{2{\rm cos}\, \theta_W}\frac{-i}{(p_1/2+p_3)^2+i\epsilon}\bar{u}(p_2)(ig_s \gamma_{\mu}T^a)  \nonumber \\
&& \cdot \frac{i}{\slashed{p}_0-\slashed{p}_1/2-m_Q+i \epsilon} \epsilon_{\nu}(p_0)\gamma^{\nu}(V_Q-A_Q\gamma_5)  \nonumber \\
&&.\Pi_1 \Lambda_1 (ig_s \gamma^{\mu}T^a) v(p_3).
\end{eqnarray}
Here, $V_Q$ and $A_Q$ are vector and axial electroweak couplings, respectively. More explicitly, $V_c=\frac{1}{2}-\frac{4}{3}\,{\rm sin}^2\theta_W$, $A_c=\frac{1}{2}$, $V_b=-\frac{1}{2}+\frac{2}{3}\,{\rm sin}^2\theta_W$ and $A_b=-\frac{1}{2}$. $\Pi_1$ is the projector for $S$-wave spin-singlet state
\begin{eqnarray}
\Pi_1=\frac{1}{(2\,m_Q)^{3/2}}(\slashed{p}_1/2-m_Q)\gamma_5(\slashed{p}_1/2+m_Q),
\end{eqnarray}
and $\Lambda_1$ is the color projector for color-singlet state
\begin{eqnarray}
\Lambda_1=\frac{\textbf{1}}{\sqrt{3}},
\end{eqnarray}
where $\textbf{1}$ is the unit matrix of the $SU(3)_c$ group.

With these amplitudes, the LO decay width for the $(Q\bar{Q})[^1S_0^{[1]}]$ pair can be calculated through
\begin{eqnarray}
d\Gamma_{\rm LO}^{(Q\bar{Q})[^1S_0^{[1]}]}=\frac{1}{3}\frac{1}{2m_{_Z}}\sum \vert M_{\rm LO} \vert^2 d\Phi_3,
\end{eqnarray}
where $\sum$ denotes the sum over the spin and color states of initial and final particles. $d\Phi_3$ is the differential phase space for the three-body final state, and
\begin{eqnarray}
d\Phi_3=(2\pi)^d\delta^d\left(p_0-\sum_{f=1}^3 p_f\right)\prod_{f=1}^3 \frac{d^{d-1} \textbf{p}_f}{(2\pi)^{d-1} 2E_f},\label{dphi3}
\end{eqnarray}
where $d$ is the number of the space-time dimensions. With these formulas, the LO decay width for $Z \to (Q\bar{Q})[^1S_0^{[1]}]+Q+\bar{Q}$ can be calculated directly.

\section{NLO corrections}
\label{secNLO}

At the NLO level, the virtual and real corrections need to be calculated. There are ultraviolet (UV) and infrared (IR) divergences in virtual correction, and IR divergence in real correction. The conventional dimensional regularization with $d=4-2\epsilon$ is employed to regularize both UV and IR divergences throughout this paper. In dimensional regularization, the $\gamma_5$ problem is notorious, and we adopt a practical prescription proposed in Ref.\cite{Korner:1991sx}. In the following subsections, we explain our main steps in calculating the virtual and real corrections.

\subsection{Virtual NLO correction}

\begin{figure}[htbp]
\includegraphics[width=0.45\textwidth]{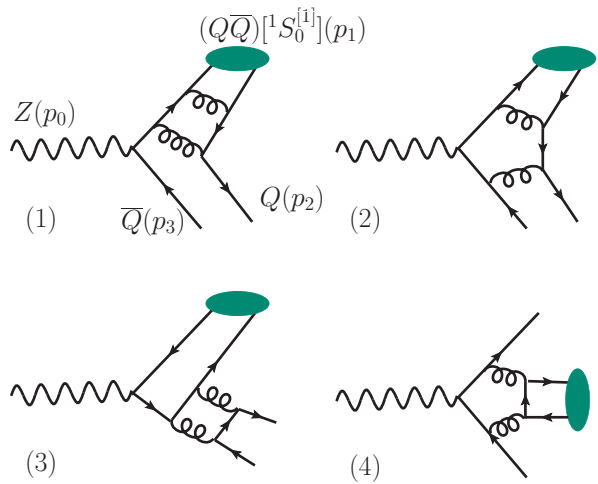}
\caption{Four typical one-loop Feynman diagrams for $Z \to (Q\bar{Q})[^1S_0^{[1]}]+Q+\bar{Q}$.
} \label{feyvir}
\end{figure}

The virtual correction at the NLO level comes from the interference of the one-loop Feynman diagrams and the LO Feynman diagrams. Four typical one-loop Feynman diagrams are shown in Fig.\ref{feyvir}. It is noted that, compared to the $J/\psi$ case \cite{JXWang}, there are new type Feynman diagrams, in which the $(Q\bar{Q})[^1S_0^{[1]}]$ pair is produced from two virtual gluons, need to be calculated in the $\eta_Q$ case. These new type Feynman diagrams do not contribute to the $J/\psi$ case. One of the new type diagrams is shown by the fourth diagram in Fig.\ref{feyvir}.

The virtual correction to the decay width of the process $Z \to (Q\bar{Q})[^1S_0^{[1]}]+Q+\bar{Q}$ can be calculated through
\begin{equation}
d\Gamma_{\rm Virtual}^{(Q\bar{Q})[^1S_0^{[1]}]}=\frac{1}{3}\frac{1}{2m_{_Z}}\sum 2 {\rm Re}\left(M^*_{\rm LO} M_{\rm Virtual} \right)d\Phi_3,
\end{equation}
where $M_{\rm Virtual}$ is the amplitude for the virtual correction, and $d\Phi_3$ is the differential three-body phase space, which has been presented in Eq.(\ref{dphi3}).

In order to take the lowest-order nonrelativistic approximation, we need to expand the amplitude in $q$ (the relative momentum between the quark and antiquark in the $(Q\bar{Q})[^1S_0^{[1]}]$ pair). In the actual calculation, we expand the amplitude in $q$ (it is equivalent to taking $q=0$ here) before performing the loop integration. In the language of method of region \cite{region}, this amounts to directly calculating the contributions from the hard region. The Coulomb divergence, which is power IR divergence, will vanish in the calculation under dimensional regularization.

There are UV and IR divergences in the loop integrals. The IR divergences from the virtual correction will be canceled by the IR divergences from the real correction. The UV divergences need to be removed through renormalization. The renormalization scheme is taken as follows: For the renormalization of the heavy quark field, the heavy quark mass and the gluon field, the on-mass-shell (OS) scheme is adopted, while for the renormalization of the strong coupling constant, the modified minimal subtraction ($\overline{\rm MS}$) scheme is adopted. With this renormalization scheme, the quantities $\delta Z_i \equiv Z_i-1$ can be derived \cite{Klasen:2004tz}
\begin{eqnarray}
 \delta Z^{\rm OS}_{2,Q}&=&-C_F \frac{\alpha_s}{4\pi}\left[\frac{1}{\epsilon_{UV}}+ \frac{2}{\epsilon_{IR}}-3~\gamma_E+3~ {\rm ln}\frac{4\pi \mu_R^2}{m_Q^2}+4\right], \nonumber\\
\delta Z^{\rm OS}_{m,Q}&=&-3~C_F \frac{\alpha_s}{4\pi}\left[\frac{1}{\epsilon_{UV}}- \gamma_E+
 {\rm ln}\frac{4\pi \mu_R^2}{m_Q^2}+\frac{4}{3}\right],\nonumber\\
 \delta Z^{\rm OS}_3&=&\frac{\alpha_s}{4\pi}\left[(\beta'_0-2C_A) \left(\frac{1}{\epsilon_{UV}}-\frac{1}{\epsilon_{IR}}\right) \right. \nonumber\\
 &&\left.-\frac{4}{3}T_F \sum_Q \left(\frac{1}{\epsilon_{UV}}-\gamma_E + {\rm ln}\frac{4\pi \mu_R^2}{m_c^2}\right)\right], \nonumber\\
 \delta Z^{\overline{\rm MS}}_g&=&- \frac{\beta_0}{2}\frac{\alpha_s}{4\pi}\left[\frac{1}{\epsilon_{UV}}- \gamma_E+ {\rm ln}~(4\pi) \right],
\end{eqnarray}
where $\mu_R$ is the renormalization scale, $\gamma_E$ is the Euler constant. $\beta_0=11C_A/3-4T_F n_f/3$ is the one-loop coefficient of the QCD $\beta$ function, in which $n_f$ is the number of active quark flavors. $\beta'_0=11C_A/3-4T_F n_{lf}/3$ and $n_{lf}=3$ is the number of light-quark flavors. When $ \mu_R \in [m_c,m_b)$, we consider the charm-quark loop in the gluon self-energy but neglect the bottom-quark and top-quark loops in the gluon self-energy, i.e. $n_f=n_{lf}+1=4$; when $ \mu_R \in [m_b,m_t)$, we consider the charm-quark and bottom-quark loops in the gluon self-energy but neglect the top-quark loop in the gluon self-energy, i.e. $n_f=n_{lf}+2=5$. For $SU(3)_c$ group, $C_A=3$, $C_F=4/3$ and $T_F=1/2$.

In the calculations, the package FeynArts \cite{feynarts} is employed to generate Feynman diagrams and amplitudes, the package FeynCalc \cite{feyncalc1,feyncalc2} is employed to carry out the color and Dirac traces, the packages \$Apart \cite{apart} and FIRE \cite{fire} are employed to conduct partial fraction and integration-by-parts (IBP) reduction. After the IBP reduction, all one-loop integrals are reduced into master integrals, and the master integrals are calculated by the package LoopTools \cite{looptools} numerically. The final phase-space integrations are calculated with the help of the package Vegas \cite{vegas}.

\subsection{Real NLO correction}

\begin{figure}[htbp]
\includegraphics[width=0.45\textwidth]{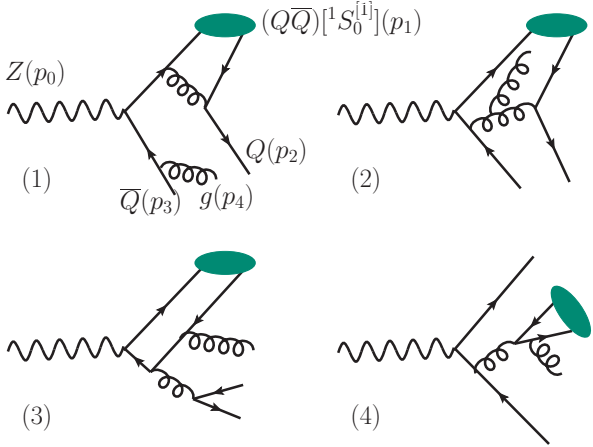}
\caption{Four typical real-correction Feynman diagrams for the decay process, $Z \to (Q\bar{Q})[^1S_0^{[1]}]+Q+\bar{Q}$.
 } \label{feyreal}
\end{figure}

The real correction to the process $Z \to (Q\bar{Q})[^1S_0^{[1]}]+Q+\bar{Q}$ comes from the process $Z(p_0) \to (Q\bar{Q})[^1S_0^{[1]}](p_1)+Q(p_2)+\bar{Q}(p_3)+g(p_4)$. Four typical Feynman diagrams are shown in Fig.\ref{feyreal}. Compared to the $J/\psi$ case, we need to deal with new type Feynman diagrams, in which the $(Q\bar{Q})[^1S_0^{[1]}]$ pair is produced from the gluon fragmentation, such as the fourth diagram of Fig.\ref{feyreal}.

Using these Feynman diagrams, the amplitude ($M_{\rm Real}$) for the real correction can be written down directly. Then the differential decay width for the real correction can be calculated through
\begin{eqnarray}
d\Gamma_{\rm Real}^{(Q\bar{Q})[^1S_0^{[1]}]}=\frac{1}{3}\frac{1}{2m_{_Z}}\sum \vert M_{\rm Real} \vert^2 d\Phi_4,
\end{eqnarray}
where $d\Phi_4$ is the differential four-body phase space,
\begin{eqnarray}
d\Phi_4=(2\pi)^d\delta^d\left(p_0-\sum_{f=1}^4 p_f\right)\prod_{f=1}^4 \frac{d^{d-1} \textbf{p}_f}{(2\pi)^{d-1} 2E_f}.
\end{eqnarray}

There are IR divergences in the real correction, which come from the phase-space integration over the region where the momentum of the final gluon is close to zero. We employ the two-cutoff phase-space slicing method \cite{twocutoff} to isolate the IR divergences in the real correction. Due to the fact that there is no collinear divergence in the present process, we only need to introduce one cutoff parameter $\delta_s$. Then the phase space for the real correction is divided into two regions: The soft region with $E_4 \leq m_{_Z} \delta_s/2$ and the hard region with $E_4 > m_{_Z} \delta_s/2$. Here, we define the gluon energy $E_4$ in the rest frame of the initial $Z$ boson. More explicitly, the real correction can be divided into two parts
\begin{eqnarray}
d\Gamma_{\rm Real}^{(Q\bar{Q})[^1S_0^{[1]}]}=d\Gamma_{\rm S}^{(Q\bar{Q})[^1S_0^{[1]}]}+d\Gamma_{\rm H}^{(Q\bar{Q})[^1S_0^{[1]}]},
\end{eqnarray}
where
\begin{eqnarray}
d\Gamma_{\rm S}^{(Q\bar{Q})[^1S_0^{[1]}]}=\frac{1}{3}\frac{1}{2m_{_Z}}\sum \vert M_{\rm Real} \vert^2 d\Phi_4 \vert_{E_4 \leq m_{_Z} \delta_s/2},
\label{eq.gamma-soft}
\end{eqnarray}
and
\begin{eqnarray}
d\Gamma_{\rm H}^{(Q\bar{Q})[^1S_0^{[1]}]}=\frac{1}{3}\frac{1}{2m_{_Z}}\sum \vert M_{\rm Real} \vert^2 d\Phi_4 \vert_{E_4 > m_{_Z} \delta_s/2}.
\end{eqnarray}

Applying the eikonal approximation to the amplitude in the soft region \cite{twocutoff,Bassetto:1983mvz}, we obtain
\begin{eqnarray}
&&\sum \vert M_{\rm Real} \vert^2 \vert_{E_4 \leq m_{_Z} \delta_s/2} \nonumber \\
&&= 4\pi \alpha_s C_F \mu_R^{2\epsilon}\bigg[ -\frac{p_2^2}{(p_2\cdot p_4)^2}+\frac{2p_2\cdot p_3}{(p_2\cdot p_4)(p_3 \cdot p_4)} \nonumber \\
&&~~~ -\frac{p_3^2}{(p_3\cdot p_4)^2}\bigg]\sum \vert M_{\rm LO} \vert^2.
\label{eq.amp-soft}
\end{eqnarray}
Up to ${\cal O}(\delta_s)$ corrections, the differential phase space for the soft region can be factorized as \cite{twocutoff}
\begin{eqnarray}
d\Phi_4 \vert_{E_4 \leq m_{_Z} \delta_s/2}=d\Phi_3 \frac{d^{d-1} \textbf{p}_4}{(2\pi)^{d-1} 2E_4}\vert_{E_4 \leq m_{_Z} \delta_s/2},
\label{eq.phs-soft}
\end{eqnarray}
where $d\Phi_3$ denotes the differential three-body phase space without emitting a gluon, whose expression has been shown in Eq.(\ref{dphi3}).

Inserting Eqs.(\ref{eq.amp-soft}) and (\ref{eq.phs-soft}) into Eq.(\ref{eq.gamma-soft}), and carrying out the integration over $p_4$ \cite{Denner:1991kt,Beenakker:2002nc}, we obtain
\begin{eqnarray}
&&d\Gamma^{(Q\bar{Q})[^1S_0^{[1]}]}_{\rm S}\nonumber\\
&&=d\Gamma^{(Q\bar{Q})[^1S_0^{[1]}]}_{\rm LO}\left[\frac{C_F \alpha_s \Gamma(1+\epsilon)}{\pi}\left(\frac{4\pi\mu_R^2}{m_{_Z}^2}\right)^{\epsilon}\right] \left\{\left( \frac{1}{\epsilon}-{\rm ln}\,\delta_s^2 \right) \right. \nonumber\\
&&~~~\times \left(1-\frac{\kappa\, p_2 \cdot p_3}{( \kappa^2-1)m_Q^2}{\rm ln}\,\kappa^2\right)+\frac{1}{2\beta_2}{\rm ln}\left(\frac{1+\beta_2}{1-\beta_2}\right)\nonumber\\
&&~~~+\frac{1}{2\beta_3}{\rm ln}\left(\frac{1+\beta_3}{1-\beta_3} \right) +\frac{2\,\kappa\, p_2 \cdot p_3}{( \kappa^2-1)m_Q^2}\left[\frac{1}{4}{\rm ln}^2\frac{u^0-\vert \textbf{u} \vert}{u^0+\vert \textbf{u}\vert} \right. \nonumber \\
&&~~~ \left.\left.\left.+{\rm Li}_2\left( 1-\frac{u^0+\vert \textbf{u} \vert}{v}\right)  +{\rm Li}_2\left( 1-\frac{u^0-\vert \textbf{u} \vert}{v}\right)\right]\right\vert^{u=\kappa\, p_2}_{u=p_3}\right\},\nonumber \\
\end{eqnarray}
where
\begin{eqnarray}
\beta_2&=&\sqrt{1-m_Q^2/E_2^2},\nonumber \\
\beta_3&=&\sqrt{1-m_Q^2/E_3^2},\nonumber \\
v&=& \frac{(\kappa^2-1)m_Q^2}{2(\kappa\, E_2-E_3)},\nonumber \\
\kappa&=& \frac{p_2\cdot p_3+\sqrt{(p_2 \cdot p_3)^2-m_Q^4}}{m_Q^2}, \nonumber
\end{eqnarray}
where $E_2$ and $E_3$ are also defined in the rest frame of the initial $Z$ boson.

Due to the constraint $E_4 > m_{_Z} \delta_s/2$ for the hard region, the contribution from the hard region is finite, then $\Gamma_{\rm H}^{(Q\bar{Q})[^1S_0^{[1]}]}$ can be numerically calculated in four dimensions. The real correction can be obtained by summing the contributions from the hard and soft regions easily. Both the contributions from the soft and hard regions are separately dependent on the cutoff parameter $\delta_s$, while the sum of these two contributions should be independent to the choice of $\delta_s$ ($\delta_s$ should be small enough.). Verifying this $\delta_s$ independence is an important test of the correctness of the calculation. We have checked the $\delta_s$ independence, and have found that the results are independent of $\delta_s$ within the error of the numerical integration when $\delta_s$ varies from $10^{-5}$ to $10^{-7}$.

The net NLO corrections can be obtained through summing the virtual and real corrections. After summing the virtual and real corrections, the UV and IR divergences are exactly canceled, and the finite results are obtained. The decay width $d\Gamma_{Z\to \eta_Q+Q\bar{Q}X}$ can be obtained from $d\Gamma_{Z\to (Q\bar{Q})[^1S_0^{[1]}]+Q\bar{Q}X}$ by multiplying a factor $\langle {\cal O}^{\eta_Q}(^1S_0^{[1]}) \rangle/\langle {\cal O}^{(Q\bar{Q})[^1S_0^{[1]}]}(^1S_0^{[1]}) \rangle \approx \vert R_{\eta_Q}(0)\vert^2/(4\pi)$, where $R_{\eta_Q}(0)$ is $\eta_Q$ radial wave function at the origin, which can be calculated by using the potential model \cite{pot}.

\section{Numerical results and discussions}
\label{secNumer}

The input parameters for the numerical calculation are taken as follows \cite{Workman:2022ynf}:
\begin{eqnarray}
&& m_c=1.67\pm 0.07\,{\rm GeV},\; m_b=4.78\pm 0.06\,{\rm GeV},\nonumber \\
&&  m_{_Z}=91.1876\,{\rm GeV},\;{\rm sin}^2\theta_W=0.231,\alpha=1/128,
\label{eq.parameter}
\end{eqnarray}
where $m_c$ and $m_b$ are the pole masses, $\alpha$ is the electromagnetic coupling constant at $m_{_Z}$. For the running strong coupling constant, we use the two-loop formula
\begin{eqnarray}
\alpha_s(\mu_R)=\frac{4\pi}{\beta_0{\rm ln}(\mu_R^2/\Lambda^2_{QCD})}\left[ 1-\frac{\beta_1{\rm ln}\,{\rm ln}(\mu_R^2/\Lambda^2_{QCD})}{\beta_0^2\,{\rm ln}(\mu_R^2/\Lambda^2_{QCD})}\right],\nonumber \\
\label{eq.alphas}
\end{eqnarray}
where $\beta_1=34\,C_A^2/3-4\,T_F \,C_F n_f-20\,T_F\, C_A n_f/3$ is the two-loop coefficient of the QCD $\beta$ function. According to $\alpha_s(m_{_Z})=0.118$~\cite{Workman:2022ynf}, we obtain $\Lambda^{n_f=5}_{\rm QCD}=0.226\,{\rm GeV}$ and $\Lambda^{n_f=4}_{\rm QCD}=0.328\,{\rm GeV}$. With the values for $\Lambda_{\rm QCD}$, the strong coupling constant at any scale can be directly calculated through Eq.(\ref{eq.alphas}).
For the radial wave functions at the origin, we adopt the values from the potential-model calculation \cite{pot}, i.e.,
\begin{eqnarray}
\vert R_{\eta_c}(0)\vert^2=0.810\,{\rm GeV}^{3},\vert R_{\eta_b}(0)\vert^2=6.477\,{\rm GeV}^{3}.
\end{eqnarray}

\subsection{Integrated decay widths}

In this subsection, we give the integrated decay widths for the decay channel $Z\to \eta_Q +Q+\bar{Q}+X$ up to the NLO level.

\begin{table}[h]
\begin{tabular}{c c c c c}
\hline
 $\mu_R$ & $\alpha_s(\mu_R)$   & ~~$\Gamma_{\rm LO}$~~ & ~~$\Gamma_{\rm NLO}$~~ & ~~$K$~~ \\
\hline
$2m_c$  & 0.245 & 62.7 & 84.3  & 1.34  \\
$m_{_Z}$ & 0.118  & 14.5 & 30.8  & 2.12 \\
\hline
\end{tabular}
\caption{The decay width (in unit: keV) of $Z\to \eta_c+c+\bar{c}+X$ up to the NLO level under two different choices of $\mu_R$, where the input charm quark mass is taken as the central value (i.e. $m_c=1.67\, {\rm GeV}$) and the K factor is defined as $K=\Gamma_{\rm NLO}/\Gamma_{\rm LO}$.}
\label{tb.widthc}
\end{table}
\begin{table}[h]
\begin{tabular}{c c c c c}
\hline
 $\mu_R$ & $\alpha_s(\mu_R)$ & ~~$\Gamma_{\rm LO}$~~ & ~~$\Gamma_{\rm NLO}$~~ & ~~$K$~~ \\
\hline
$2m_b$   & 0.180 & 10.8 & 13.8  & 1.28  \\
$m_{_Z}$ & 0.118 & 4.65 & 8.49  & 1.83 \\
\hline
\end{tabular}
\caption{The decay width (in unit: keV) of $Z\to \eta_b+b+\bar{b}+X$ up to the NLO level under two different choices of $\mu_R$, where the input bottom quark mass is taken as the central value (i.e. $m_b=4.78\, {\rm GeV}$) and the K factor is defined as $K=\Gamma_{\rm NLO}/\Gamma_{\rm LO}$.}
\label{tb.widthb}
\end{table}

\begin{figure}[htbp]
\includegraphics[width=0.45\textwidth]{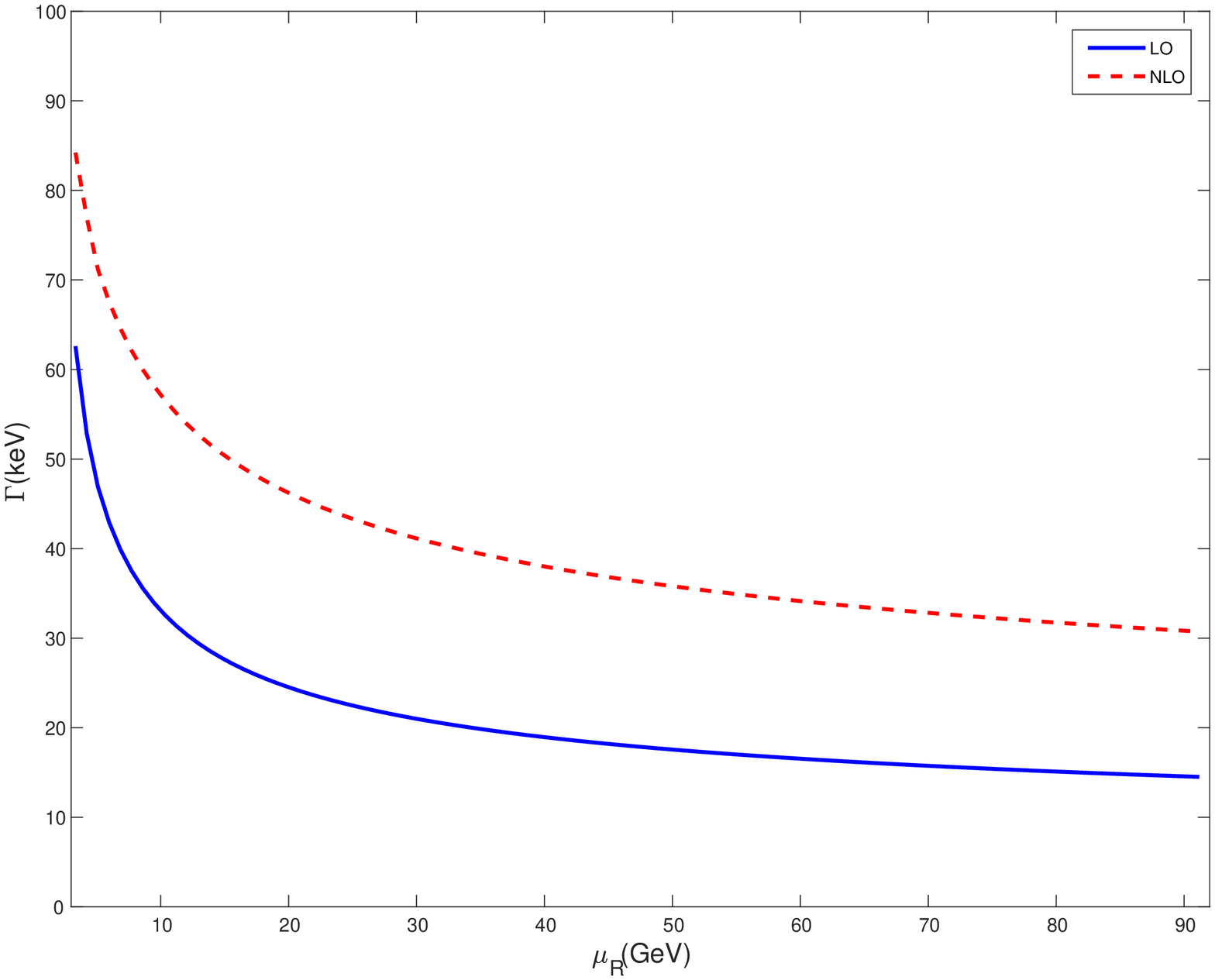}
\caption{The LO and NLO decay widths for $Z \to \eta_c+c+\bar{c}+X$ as functions of the renormalization scale $\mu_R$.}
\label{gammamurc}
\end{figure}

\begin{figure}[htbp]
\includegraphics[width=0.45\textwidth]{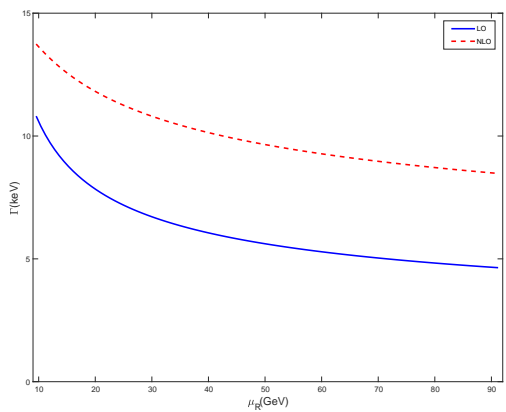}
\caption{The LO and NLO decay widths for $Z \to \eta_b+b+\bar{b}+X$ as functions of the renormalization scale $\mu_R$.}
\label{gammamurb}
\end{figure}

In order to have a glance on the size of the NLO corrections, we first present the decay widths when the input quark masses are taken as their central values (i.e., $m_c=1.67\, {\rm GeV}$ and $m_b=4.78\, {\rm GeV}$), an analysis of the uncertainties of these decay widths will be presented later. The decay widths of $Z\to \eta_Q +Q+\bar{Q}+X$ up to the NLO level are given in Tables \ref{tb.widthc} and \ref{tb.widthb}, where $\Gamma_{\rm NLO}$ denotes the sum of the LO contribution and the NLO corrections. The renormalization scales are set as two energy scales involved in the processes, i.e., $2m_Q$ and $m_{_Z}$. Tables \ref{tb.widthc} and \ref{tb.widthb} show that the NLO corrections contribute significantly to the decay widths in both cases. The NLO corrections increase the decay width of $Z\to \eta_c+c+\bar{c}+X$ by $\sim 34\%$ for $\mu_R=2 m_c$ and $\sim 112\%$ for $\mu_R=m_{_Z}$; and they increase the decay width of $Z\to \eta_b+b+\bar{b}+X$ by $\sim 28\%$ for $\mu_R=2 m_b$ and $\sim 83\%$ for $\mu_R=m_{_Z}$.

In Figs.\ref{gammamurc} and \ref{gammamurb}, the dependence of the decay widths on the renormalization scale is shown. After including the NLO corrections, the dependence of the decay widths on the renormalization scale is weakened. For $Z\to \eta_c+c+\bar{c}+X (Z\to \eta_b+b+\bar{b}+X)$, the decay width decreases by $77\%$($57\%$) at LO, and by $63\%$($39\%$) at NLO when $\mu_R$ varies from $2m_Q$ to $m_{_Z}$. However, this dependence is still strong even including the NLO corrections.

Now, let us estimate the theoretical uncertainties for these decay widths. The main uncertainty sources include the renormalization scale, the heavy quark masses and the radial wave functions at the origin \footnote{The Monte Carlo numerical integration in the calculation would lead to an error. However, the error of the numerical integration is on the order of $10^{-2}\,{\rm keV}$ for the $\eta_c$ case, and $10^{-3}\,{\rm keV}$ for the $\eta_b$ case in our calculation. Thus, the error from the numerical integration is negligible.}. For the uncertainties caused by the renormalization scale, we estimate them by varying the renormalization scale between two physical energy scales involved in the processes, i.e. $2m_Q$ and $m_{_Z}$. Furthermore, we take the average values of the decay widths under the two choices of the renormalization scale as their central values. For the uncertainties caused by the heavy quark masses, we estimate them by varying the heavy quark masses in the ranges given in Eq.(\ref{eq.parameter}), i.e., $m_c=1.67\pm 0.07\,{\rm GeV}$ and $m_b=4.78\pm 0.06\,{\rm GeV}$. For the radial wave functions at the origin, the authors of Ref.\cite{pot} did not give an error estimate. Since the potential used in Ref.\cite{pot} does not include the spin effect, the wave functions calculated in this way are accurate up to corrections of relative order $v_Q^2$. Therefore, we estimate the uncertainties by attaching an error of $30\%$ of the central value for $\eta_c$, and $10\%$ of the central value for $\eta_b$. More explicitly, we take $\vert R_{\eta_c}(0)\vert^2=0.810\pm 0.243\,{\rm GeV}^{3}$ and $\vert R_{\eta_b}(0)\vert^2=6.477\pm 0.648\,{\rm GeV}^{3}$. Then we obtain
\begin{eqnarray}
\Gamma^{\rm LO}_{Z \to \eta_c+c\bar{c}X} &=& 38.6^{+24.1+6.7+11.6}_{-24.1-5.5-11.6}\,{\rm keV},\nonumber \\
\Gamma^{\rm NLO}_{Z \to \eta_c+c\bar{c}X} &=& 57.6^{+26.7+10.3+17.3}_{-26.8-8.4-17.3}\,{\rm keV},
\label{eq.uncert1}
\end{eqnarray}
and
\begin{eqnarray}
\Gamma^{\rm LO}_{Z \to \eta_b+b\bar{b}X} &=& 7.74^{+3.09+0.35+0.78}_{-3.09-0.40+0.78}\,{\rm keV},\nonumber \\
\Gamma^{\rm NLO}_{Z \to \eta_b+b\bar{b}X} &=& 11.1^{+2.7+0.5+1.2}_{-2.7-0.5-1.2}\,{\rm keV}.
\label{eq.uncert2}
\end{eqnarray}
Here, the first error is caused by the renormalization scale, the second error is caused by the heavy quark mass, and the last error is caused by the radial wave function at the origin. From Eqs.(\ref{eq.uncert1}) and (\ref{eq.uncert2}), we can see that the largest error arises from the renormalization scale uncertainty for both $\eta_c$ and $\eta_b$ cases. Furthermore, we find that although the K factors are sensitive to the renormalization scale, they are insensitive to the heavy quark mass, e.g., when we vary the charm (bottom) quark mass from $1.60{\rm GeV}$ ($4.72{\rm GeV}$) to $1.74{\rm GeV}$ ($4.84{\rm GeV}$), the K factor changes from $1.50$ ($1.43$) to $1.49$ ($1.44$) for the $\eta_c$ ($\eta_b$) case.

Adding the uncertainties in quadrature, we obtain
\begin{eqnarray}
\Gamma^{\rm LO}_{Z \to \eta_c+c\bar{c}X} &=& 38.6^{+27.6}_{-27.3}\,{\rm keV},\nonumber \\
\Gamma^{\rm NLO}_{Z \to \eta_c+c\bar{c}X} &=& 57.6^{+33.4}_{-33.0}\,{\rm keV},
\label{eq.uncert3}
\end{eqnarray}
and
\begin{eqnarray}
\Gamma^{\rm LO}_{Z \to \eta_b+b\bar{b}X} &=& 7.74^{+3.21}_{-3.21}\,{\rm keV},\nonumber \\
\Gamma^{\rm NLO}_{Z \to \eta_b+b\bar{b}X} &=& 11.1^{+3.0}_{-3.0}\,{\rm keV}.
\label{eq.uncert4}
\end{eqnarray}

\subsection{Differential decay widths}

\begin{figure}[htbp]
\includegraphics[width=0.45\textwidth]{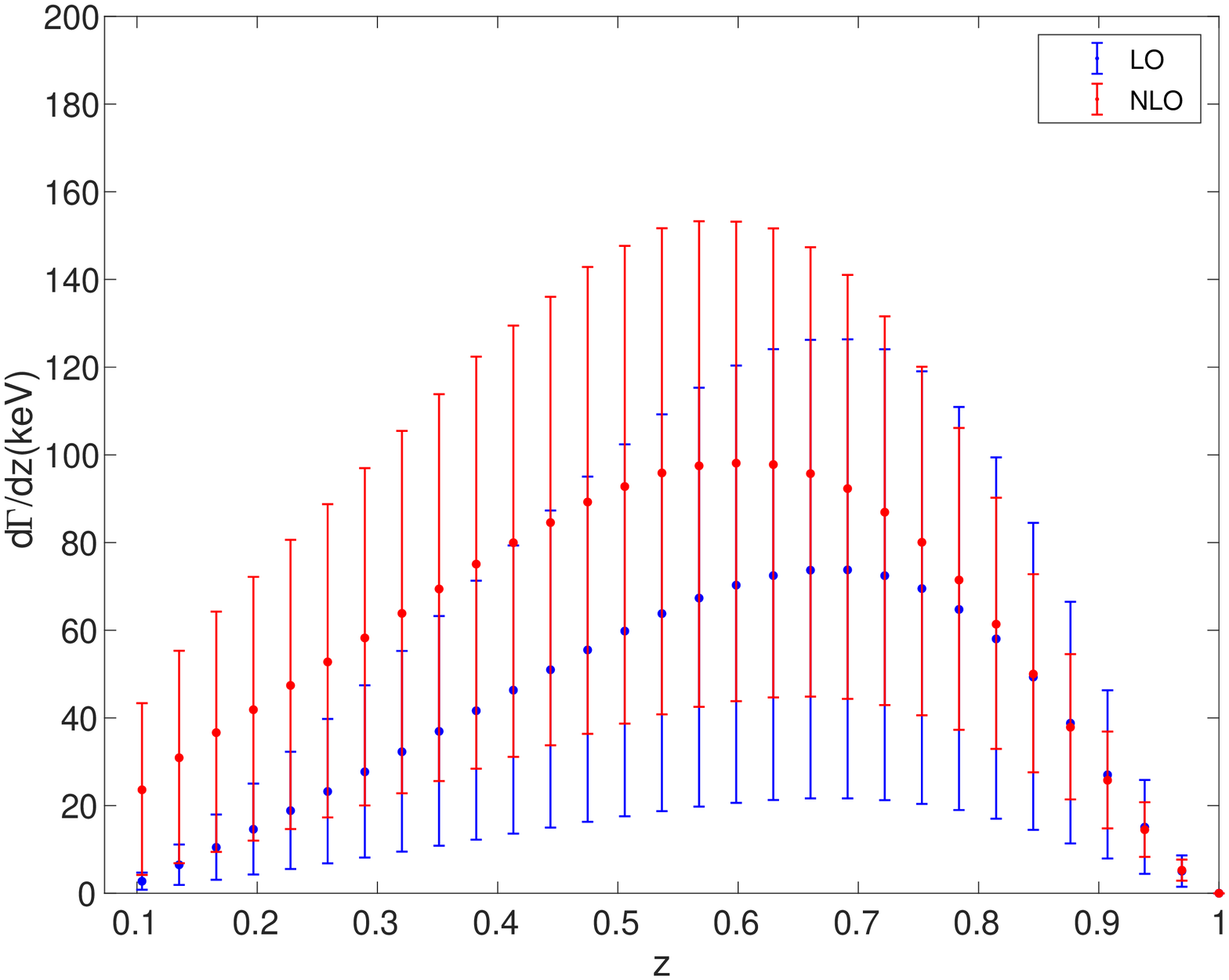}
\caption{The LO and NLO differential decay widths $d\Gamma/dz$ for $Z \to \eta_c+c+\bar{c}+X$. The error bars show the total uncertainties caused by the renormalization scale, the heavy quark mass and the wave function at the origin, and the total uncertainties are obtained by adding each uncertainty in quadrature.}
\label{gammazc}
\end{figure}

\begin{figure}[htbp]
\includegraphics[width=0.45\textwidth]{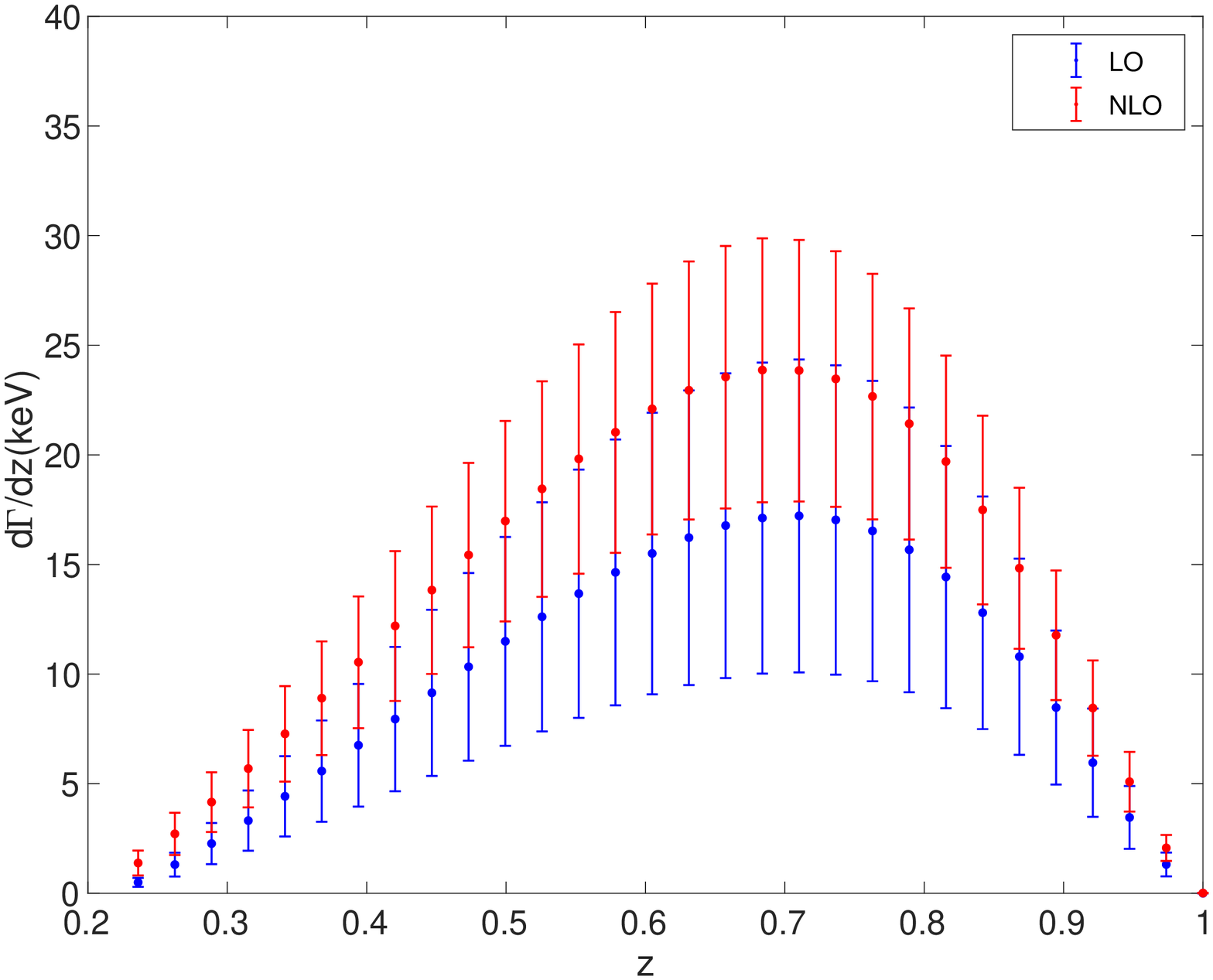}
\caption{The LO and NLO differential decay widths $d\Gamma/dz$ for $Z \to \eta_b+b+\bar{b}+X$. The error bars show the total uncertainties caused by the renormalization scale, the heavy quark mass and the wave function at the origin, and the total uncertainties are obtained by adding each uncertainty in quadrature.}
\label{gammazb}
\end{figure}

\begin{figure}[htbp]
\includegraphics[width=0.45\textwidth]{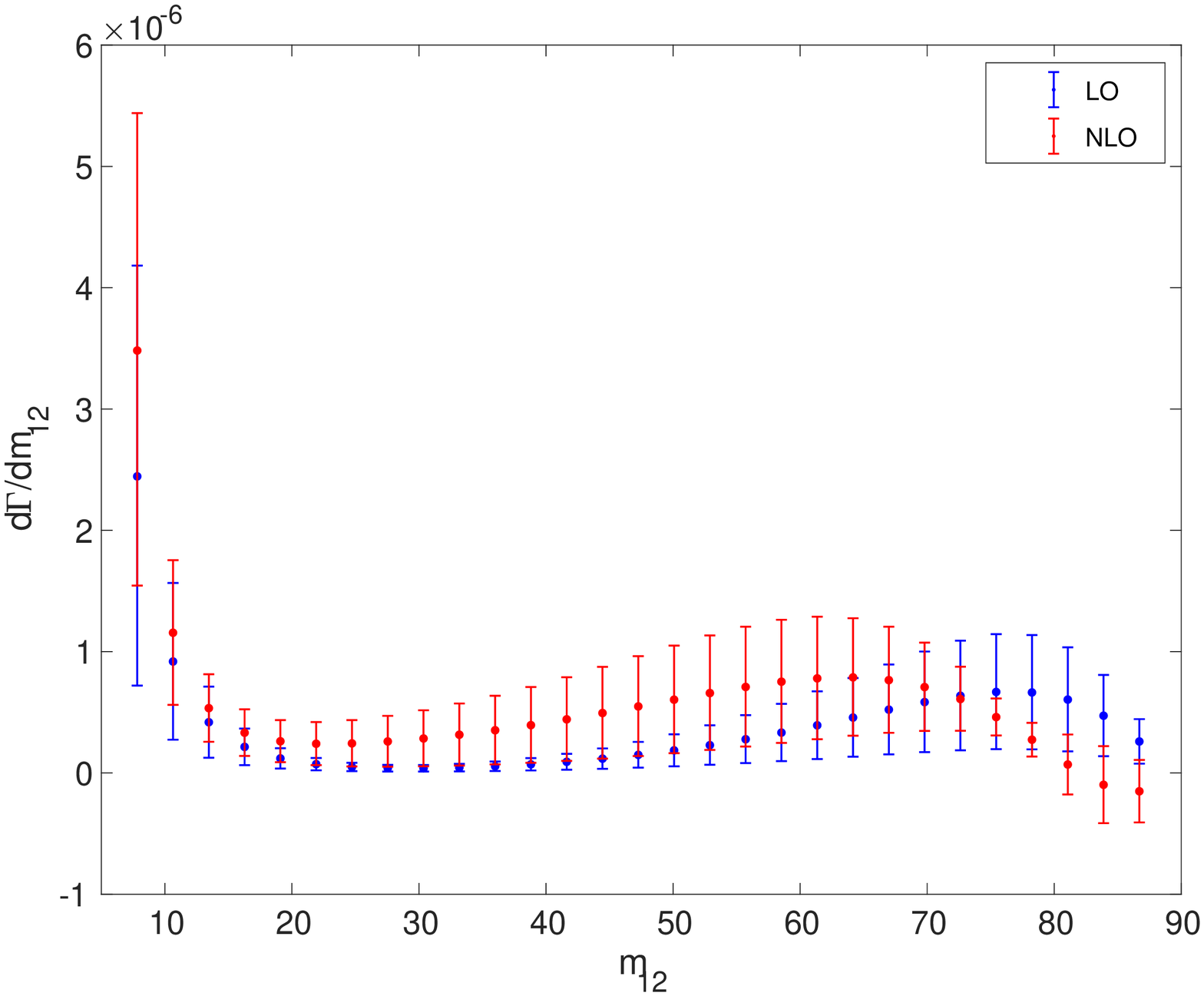}
\caption{The LO and NLO differential decay widths $d\Gamma/d m_{12}$ for $Z \to \eta_c+c+\bar{c}+X$. The error bars show the total uncertainties caused by the renormalization scale, the heavy quark mass and the wave function at the origin, and the total uncertainties are obtained by adding each uncertainty in quadrature.}
\label{gammam12c}
\end{figure}

\begin{figure}[htbp]
\includegraphics[width=0.45\textwidth]{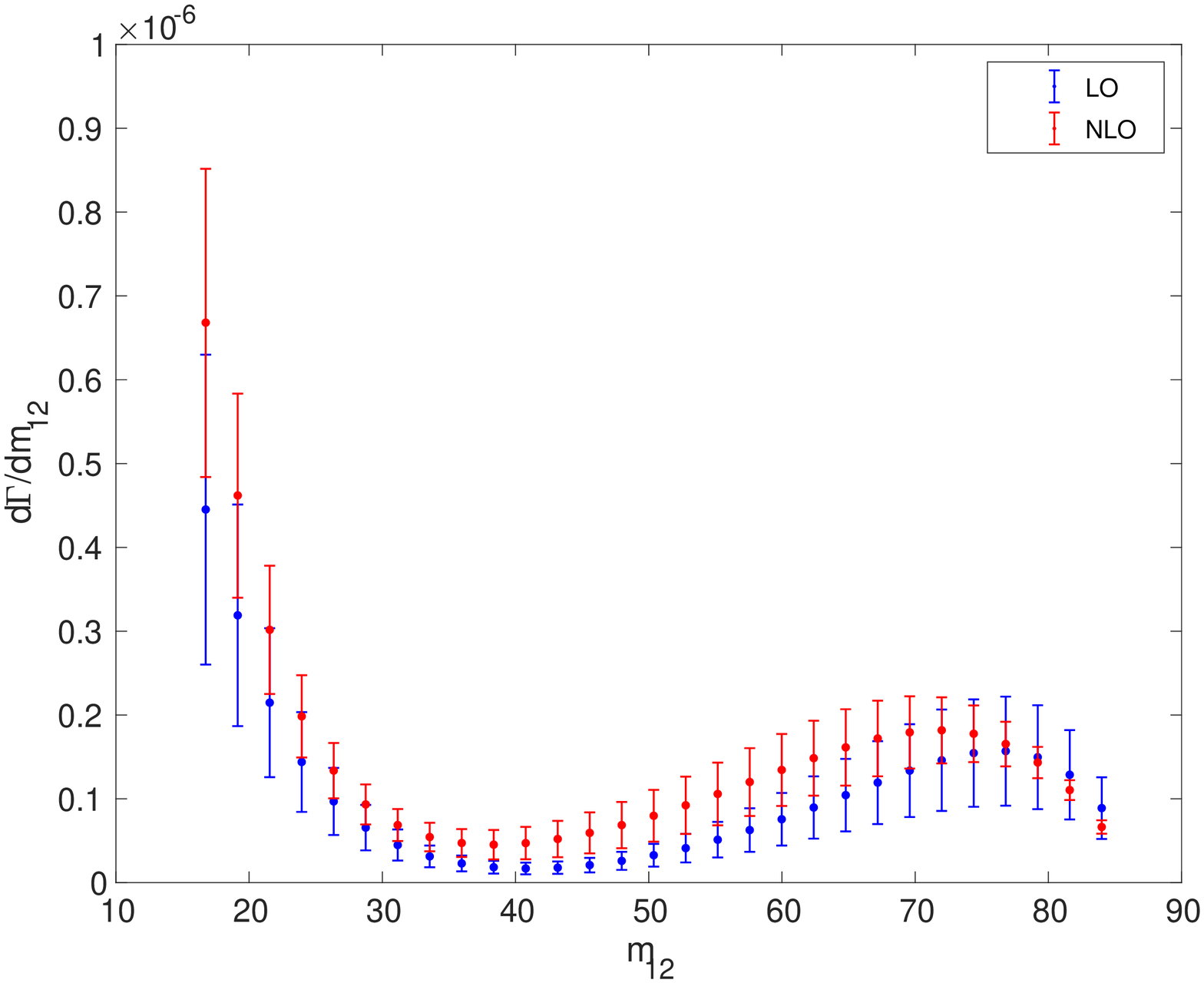}
\caption{The LO and NLO differential decay widths $d\Gamma/dm_{12}$ for $Z \to \eta_b+b+\bar{b}+X$. The error bars show the total uncertainties caused by the renormalization scale, the heavy quark mass and the wave function at the origin, and the total uncertainties are obtained by adding each uncertainty in quadrature.}
\label{gammam12b}
\end{figure}

\begin{figure}[htbp]
\includegraphics[width=0.45\textwidth]{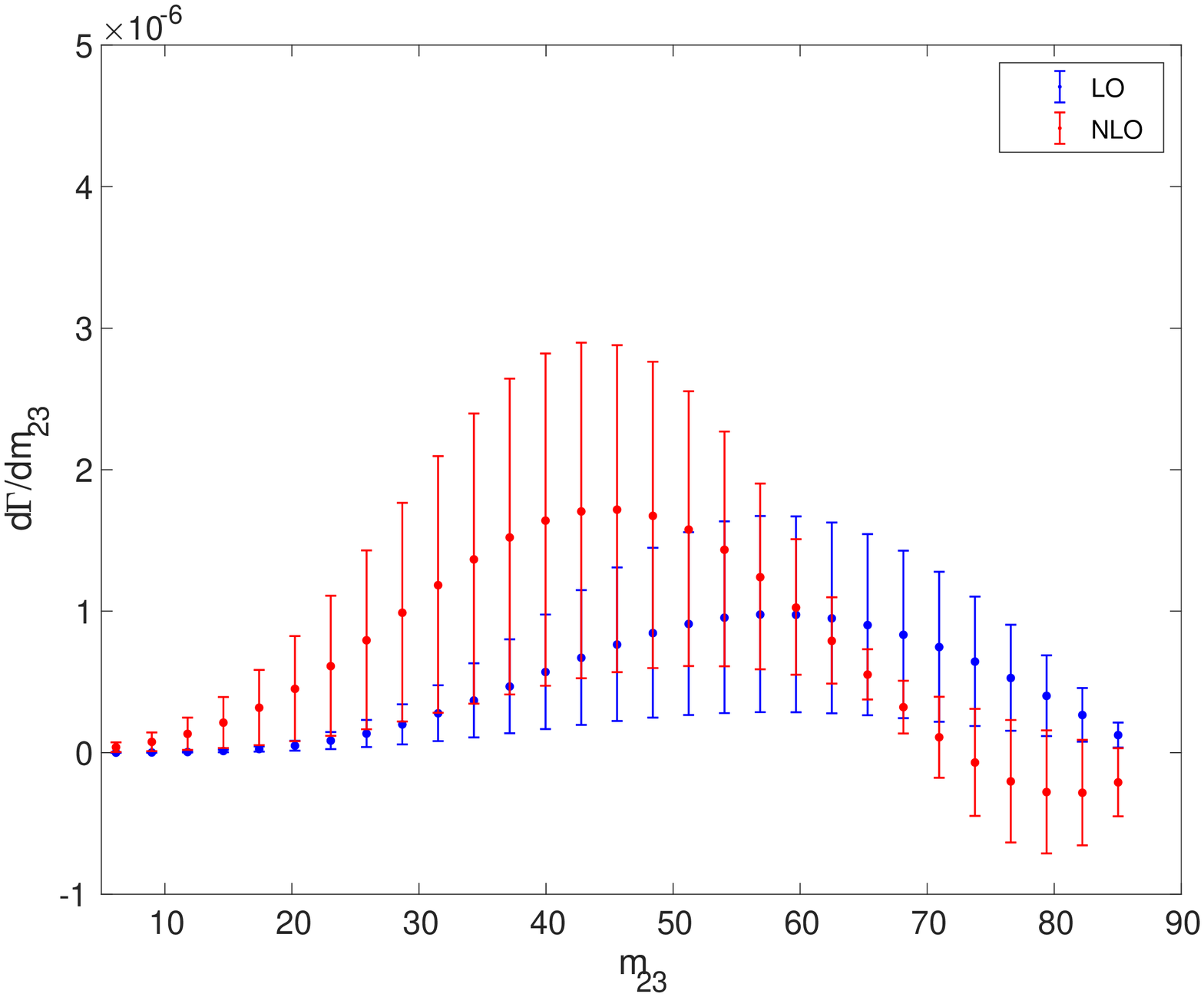}
\caption{The LO and NLO differential decay widths $d\Gamma/d m_{23}$ for $Z \to \eta_c+c+\bar{c}+X$. The error bars show the total uncertainties caused by the renormalization scale, the heavy quark mass and the wave function at the origin, and the total uncertainties are obtained by adding each uncertainty in quadrature.}
\label{gammam23c}
\end{figure}

\begin{figure}[htbp]
\includegraphics[width=0.45\textwidth]{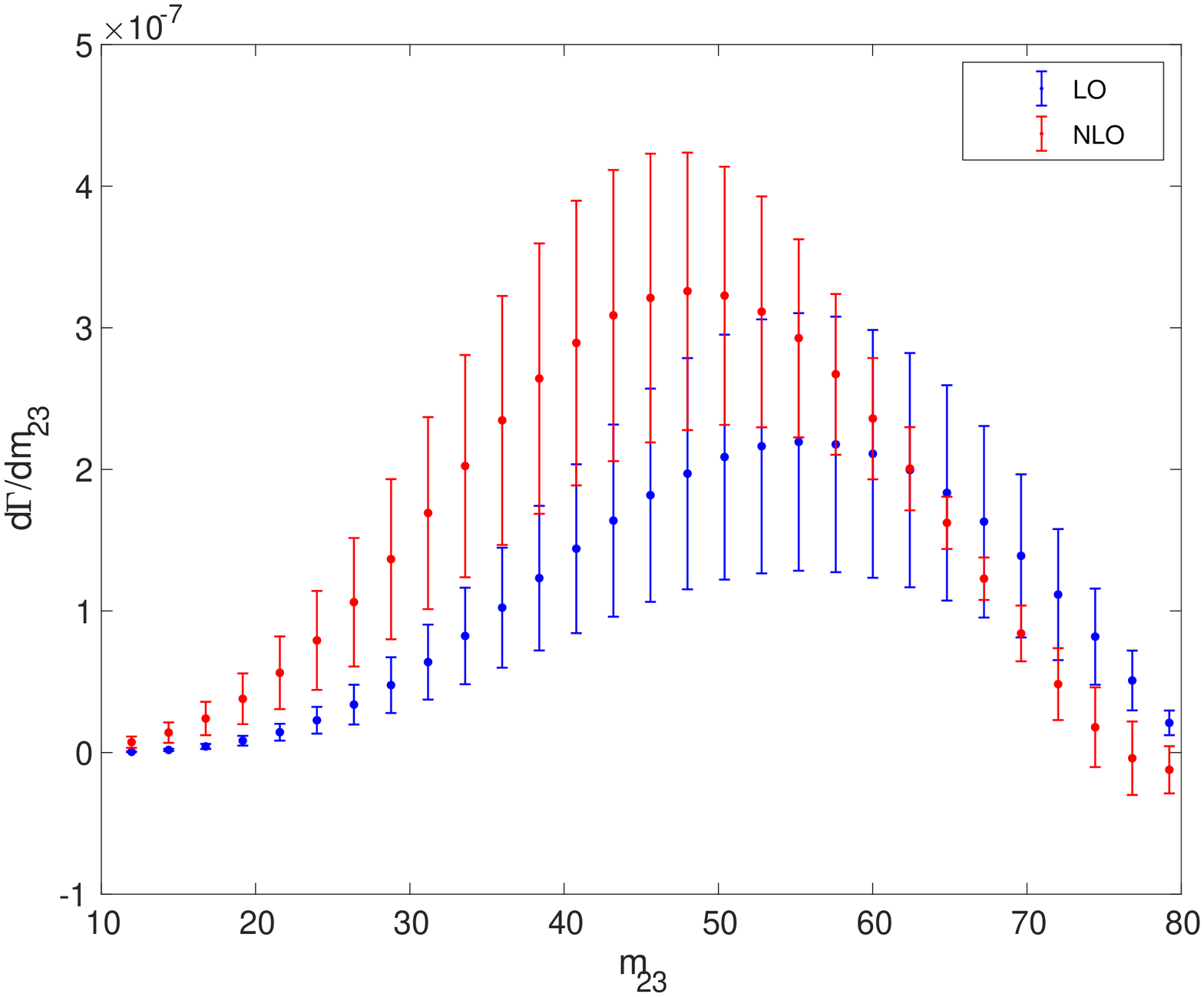}
\caption{The LO and NLO differential decay widths $d\Gamma/dm_{23}$ for $Z \to \eta_b+b+\bar{b}+X$. The error bars show the total uncertainties caused by the renormalization scale, the heavy quark mass and the wave function at the origin, and the total uncertainties are obtained by adding each uncertainty in quadrature.}
\label{gammam23b}
\end{figure}

The differential distributions contain more information than the integrated decay widths, which can be used to test the current theory. Therefore, it is interesting to see the differential distributions of the two $Z$ boson decay processes.

The energy fraction carrying by $\eta_c$ or $\eta_b$ in the processes can be defined as $z \equiv 2\,p_0 \cdot p_1/m_{_Z}^2$. The LO and NLO differential decay widths $d\Gamma/dz$ for the processes $Z\to \eta_c+c+\bar{c}+X$ and $Z\to \eta_b+b+\bar{b}+X$ are shown in Figs. \ref{gammazc} and \ref{gammazb}, respectively. Figs. \ref{gammazc} and \ref{gammazb} confirm the importance of the NLO corrections. For the $\eta_c$ production, the magnitude of $d\Gamma/dz$ is increased obviously at small and moderate $z$ values and decreased slightly at higher $z$ values. And for the $\eta_b$ production, the magnitude of $d\Gamma/dz$ is increased at all $z$ values. The uncertainties for $d\Gamma/dz$ are also shown in the figures, which are obtained by combining the uncertainties of the renormalization scale, the heavy quark mass and the wave function at the origin.

The momenta of heavy quarks in the final state can be determined experimentally using the heavy-flavor tagging technology. Therefore, the differential decay widths $d\Gamma/dm_{12}$ and $d\Gamma/dm_{23}$ can be measured experimentally, where $m_{12} \equiv \sqrt{(p_1+p_2)^2}$ and $m_{23} \equiv \sqrt{(p_2+p_3)^2}$ are invariant masses of two final-state particles. We present the LO and NLO differential decay widths $d\Gamma/dm_{12}$ and $d\Gamma/dm_{23}$ for $Z\to \eta_c +c+\bar{c}+X$ and $Z\to \eta_b +b+\bar{b}+X$ in Figs. \ref{gammam12c}, \ref{gammam12b}, \ref{gammam23c} and \ref{gammam23b}, respectively. The uncertainties for these differential decay widths are also shown in these figures. From the figures, we can see that the differential decay widths $d\Gamma/dm_{12}$ and $d\Gamma/dm_{23}$ are changed significantly after including the NLO corrections, especially for the decay $Z\to \eta_c+c+\bar{c}+X$. Moreover, it is found that the NLO differential decay widths $d\Gamma/dm_{12}$ and $d\Gamma/dm_{23}$ are negative at the large $m_{12}$ or $m_{23}$ for $Z\to \eta_c +c+\bar{c}+X$. This indicates that the NLO corrections are negative and larger than LO contribution in these phase space regions. In the boundary regions of the phase space, some large logarithmic terms often appear in the coefficients of the perturbation expansion, which may spoil the convergence of the perturbation series. In these boundary regions, it is necessary to resum these large logarithmic terms so as to obtain precise theoretical results. Because of these large logarithmic terms, if we calculate to a certain order (e.g. NLO) in these regions, we may obtain nonphysical negative results. Fortunately, in the considered processes, the absolute values of the negative contributions of these regions are not large. Thus, we believe that the differential decay widths for most regions of the phase space and the integrated decay widths, which are obtained in this paper, are reliable.

\section{Summary}
\label{secSum}

In the present paper, we have studied the decays $Z \to \eta_c+c+\bar{c}+X$ and $Z \to \eta_b+b+\bar{b}+X$ up to NLO QCD accuracy. Integrated and differential decay widths of both decay processes are obtained, and the uncertainties for them are estimated. We find that the NLO corrections to the decay widths for $Z \to \eta_c+c+\bar{c}+X$ and $Z \to \eta_b+b+\bar{b}+X$ are significant. The dependence of the decay widths on the renormalization scale is very strong although the dependence is weakened after including the NLO corrections. This brings a big uncertainty to the theoretical predictions under the conventional renormalization scale setting. The higher-order corrections can reduce the uncertainty caused by the renormalization scale. However, it is very difficult to calculate the higher-order corrections beyond the NLO for these processes at present. In the literature, the principle of maximum conformality (PMC) scale-setting approach~\cite{pmc1, pmc2, pmc4, pmc5} has been suggested to eliminate such scale uncertainty, whose key idea is to determine the correct momentum flow of the process by using the nonconformal $\beta$-terms that govern the $\alpha_s$ running behavior with the help of renormalization group equation. As a comparison, we also calculate the integrated decay widths under the PMC scale setting. Following the standard PMC procedures to the decay width of $Z \to \eta_Q+Q+\bar{Q}+X$, we obtain $\Gamma_{Z\to \eta_c +c\bar{c}X}^{\rm PMC}=95.4^{+34.1}_{-32.2}\,{\rm keV}$ and $\Gamma_{Z\to \eta_b +b\bar{b}X}^{\rm PMC}=14.7^{+1.7}_{-1.7}\,{\rm keV}$. Here, the PMC predictions of the decay width are independent to the choices of $\mu_R$, and the errors come from the uncertainties of the heavy quark masses and the wave functions at the origin.

The cross sections for $e^+e^- \to Z \to \eta_Q+Q+\bar{Q}+X$ at the $Z$ pole~\footnote{The $\gamma$-exchange contribution is negligibly small at the $Z$ pole, which can be safely neglected.} can be derived from the decay widths $\Gamma_{Z \to \eta_Q+Q\bar{Q}X}$ through the formulas derived in the Appendix A1 of Ref.\cite{Zheng:2017xgj}, i.e.,
\begin{eqnarray}
\sigma_{e^+e^-  \to \eta_Q+Q\bar{Q}X}=&& \frac{e^2(1-4{\rm sin}^2\theta_W + 8{\rm sin}^4\theta_W)}{8{\rm sin}^2\theta_W {\rm cos}^2\theta_W m_{_Z} \Gamma_Z^2}\nonumber \\
&& \times \Gamma_{Z  \to \eta_Q+Q\bar{Q}X}.
\end{eqnarray}
Then we obtain
\begin{eqnarray}
\sigma_{e^+e^-  \to \eta_c+c\bar{c}X}&=&1.37^{+0.80}_{-0.78}\,{\rm pb},  \\
\sigma_{e^+e^-  \to \eta_b+b\bar{b}X}&=&0.264^{+0.072}_{-0.071}\,{\rm pb}.
\end{eqnarray}
If the luminosity of a $Z$ factory can be up to $10^{35}{\rm cm}^{-2}{\rm s}^{-1}$~\cite{zfactory}, there are about $1.4 \times 10^{6}$ $\eta_c+c\bar{c}X$ events and $2.6 \times 10^{5}$ $\eta_b+b\bar{b}X$ events to be produced per operation year. Moreover, the background of the Z factory is clean. Therefore, the two production processes may be studied at a high luminosity $Z$ factory.

\hspace{2cm}

\noindent {\bf Acknowledgments:} This work was supported in part by the Natural Science Foundation of China under Grants No. 12005028, No. 12175025, No. 12147116 and No. 12147102, by the China Postdoctoral Science Foundation under Grant No. 2021M693743, by the Fundamental Research Funds for the Central Universities under Grant No. 2020CQJQY-Z003, and by the Chongqing Graduate Research and Innovation Foundation under Grant No. ydstd1912.

\hspace{2cm}

\end{document}